\documentclass[referee]{raa}            
\usepackage{graphicx,times}             
\usepackage{natbib}
\usepackage{amssymb,amsmath}
\usepackage{color}
\usepackage{ulem}
\usepackage[dvipsnames]{xcolor}
\RequirePackage{lineno}
\usepackage{cancel}
\usepackage[OT1]{fontenc}
\usepackage{siunitx}
\usepackage{gensymb}
\usepackage{enumitem}
\usepackage{xcolor}

\bibpunct{(}{)}{;}{a}{}{,}

\usepackage[pagebackref=true]{hyperref}
\usepackage{siunitx}

\begin{document}

  \title{ Resolution enhancement of SOHO/MDI Magnetograms 
}

   \volnopage{Vol.0 (20xx) No.0, 000--000}      
   \setcounter{page}{1}          

   \author{Ying Qin 
   \inst{1,2,3}
   \and Kai-Fan Ji 
      \inst{1,3,4}
   \and Hui Liu
      \inst{1,3,4}   
   \and Xiao-Guang Yu  
      \inst{1,3,4}
    }

   \institute{Yunan Observatories, Chinese Academy of Sciences,
   Kunming 650011,Yunan, China; {\\ jkf@ynao.ac.cn; qinying@ynao.ac.cn}\\
       \and
           University of Chinese Academy of Sciences, Beijing 100049, China\\
       \and 
           Yunnan Key Laboratory of Solar Physics and Space Science, 650216, China
        \and   
           Key Laboratory for the Structure and Evolution of Celestial Objects, Chinese Academy of Sciences, Kunming 650216,China\\ 
 {\small Received 20xx month day; accepted 20xx month day}}
  
  \abstract{Research on the solar magnetic field and its effects on solar dynamo mechanisms and space weather events has benefited from the continual improvements in instrument resolution and measurement frequency. The augmentation and assimilation of historical observational data timelines also play a significant role in understanding the patterns of solar magnetic field variation. Within the realm of astronomical data processing, super-resolution reconstruction refers to the process of using a substantial corpus of training data to learn the nonlinear mapping between low-resolution and high-resolution images, thereby achieving higher-resolution astronomical images. This paper is an application study in high-dimensional non-linear regression. Deep learning models were employed to perform super-resolution (SR) modeling on SOHO/MDI magnetograms and SDO/HMI magnetograms, thus reliably achieving resolution enhancement of full-disk SOHO/MDI magnetograms and enhancing the image resolution to obtain more detailed information. For this study, a dataset comprising 9,717 pairs of data from April 2010 to February 2011 was used as the training set, 1,332 pairs from March 2011 were used as the validation set, and 1,034 pairs from April 2011 were used as the test set. After data preprocessing, we randomly cropped 128x128 sub-images as the low resolution (LR) from the full-disk MDI magnetograms, and the corresponding 512x512 sub-images as high resolution (HR) from the HMI full-disk magnetograms for model training. The tests conducted have shown that the study successfully produced reliable 4x super-resolution reconstruction of full-disk MDI magnetograms. The MESR model's results (0.911) were highly correlated with the target HMI magnetographs as indicated by the correlation coefficient values. Furthermore, the method achieved the best PSNR, SSIM, MAE and RMSE values, indicating that the MESR model can effectively reconstruct magnetograms.
    \keywords{techniques: image processing,virtual observatory tools,techniques: image processing,Sun: magnetic fields }
}
    \authorrunning{Ying Qin et al}            
    \titlerunning{Resolution enhancement of SOHO/MDI Magnetograms}           
    \maketitle

   \section{Introduction}           
\label{sect:intro}

The solar magnetic field plays a pivotal role in solar activities. Many significant achievements and principal challenges in the field of solar physics are related to the observation and theoretical study of the solar magnetic field. The images depicting the solar magnetic field, which illustrate field strength, polarity, and distribution across the sun, provide vital evidence for scientists to explore the field distribution, connection between active regions, coronal activities, and magnetic hydrodynamic simulations(\citealt{2012The}). Thus, obtaining high-quality magnetic field data (magnetograms) is especially critical.
\\
Since the 1990s, Europe, the USA, and Japan have successively launched space satellites dedicated to observing the solar magnetic field, including the Solar Heliospheric Observatory (\citealt{1995The}), Hinode (\citealt{kosugi2008hinode}), and the Solar Dynamics Observatory (\citealt{pesnell2012solar}). Of these, the SOHO satellite represents humanity's first true space-based solar magnetic field observatory, equipped with the Michelson Doppler Imager(\citealt{domingo1995soho}), which features a 14cm aperture and a spatial resolution of $2\arcsec/pix$. However,the MDI is limited to only line-of-sight magnetic field observations, lacking vector magnetic field measurements, and it is hampered by low temporal and spatial resolutions. Consequently, scientific research based on its magnetic field data has thus far only made strides in addressing issues related to the solar activity cycle. It has not provided sufficient high-resolution observational data to explore more substantial questions such as the magnetic activities in the solar atmosphere, the intrinsic nature of the solar magnetic field, and the magnetic energy release mechanisms during solar eruptive events. For various reasons, the SOHO/MDI observation program was terminated on April 12, 2011, with the primary role in spatial vector magnetic field observation now taken over by the Helioseismic and Magnetic Imager (\citealt{schou2012design}) aboard the SDO satellite, featuring a 12.5 cm aperture and a spatial resolution of $0.5\arcsec/pix$, realizing the first spatial observations of the full-disk solar vector magnetic field. This instrument’s main mission is to investigate the mechanisms of solar activity variations, to depict and understand physical processes inside the sun, and to observe the magnetic activities in the solar atmosphere. Despite HMI being designed with a polarization selector and image stabilization system, its spatial resolution remains insufficient for observing small-scale solar structures. By contrast, the Hinode satellite, operational since 2006, has been the first to use a 50 cm optical telescope for spatial observations of solar optics and magnetic fields, achieving a spatial resolution of $0.16\arcsec/pix$ and providing high-resolution images of local solar magnetic fields.
\\
Compared to low-resolution images (SOHO/MDI), high-resolution images (SDO/HMI, Hinode/SOT) usually contain greater pixel density, richer texture details, and higher reliability. Likewise, compared to local magnetic field observations (Hinode/SOT), full-disk magnetic field data (SOHO/MDI, SDO/HMI) offer additional information for studies related to the interconnectedness of solar eruptive activities and the evolution of small-scale magnetic phenomena(\citealt{baso2018enhancing}). Integration of historical observational data over a timeline (from SOHO/MDI to SDO/HMI) is of great importance for studying the patterns of solar magnetic field changes.
\\
In recent years, an increasing number of scholars have also been attempting to apply deep learning techniques to solar magnetic field image processing. Anna Jungbluth and colleagues (\citealt{2019Single}) employed a HighRes-net based on a Multi-Frame Super-Resolution (MFSR) architecture, leveraging an arbitrary number of low-resolution (LR) images from the same view to super-resolve MDI magnetograms to the same resolution as HMI magnetograms. Sumiaya Rahman and others applied a residual attention model and progressive GAN model on 4x4 binning downsampled SDO/HMI magnetograms(\citealt{2020Super}), comparing the model-generated outcomes with corresponding Hinode/NFI magnetograms. Dou and colleagues (\citealt{dou2022super}) utilized two GANs to first transfer MDI features onto downsampled HMI, obtaining low-resolution HMI magnetograms in the same domain as MDI, and then employed paired low-resolution and high-resolution HMI magnetograms to generate visually high-quality super-resolved magnetograms. However, the network used by Jungbluth et al. was relatively basic and did not take advantage of the complete 2010 - 2011 MDI and HMI overlapping magnetograms, leading to subpar reconstruction results. Rahman et al.'s study trained the network with simulated data obtained from 2D Gaussian smoothing and bilinear interpolation downsampled, leading to a mismatch between the final testing scenario and the training scenario.
\\
As a result of the similar aperture sizes of the SOHO/MDI and SDO/HMI telescopes and early CCD technology limitations, the resolution of MDI images was only a quarter of that of HMI images (undersampling)(\citealt{liu2012comparison}). In this paper, we aim to harness the powerful capability of deep learning(\citealt{dong2015image}) to discern patterns from sample data to construct a deep learning network apt for super-resolution reconstruction of solar magnetic field images, thereby achieving a reliable 4x resolution  enhancement of the SOHO/MDI (1996~2011) solar magnetic field images from historical observational data since 1996. This will support research that requires high-resolution and homogeneous data. We utilized 20,150 sets of simultaneous observations of identical physical structures from SOHO/MDI and SDO/HMI between April 8, 2010, and April 11, 2011, as samples. Through preprocessing operations such as rotation, translation, and scaling to align the images, we randomly cropped 128x128 blocks as LR images on full-disk MDI magnetograms and corresponding 512x512 blocks as HR images on full-disk
 HMI magnetograms. We structured a model based on the Residual Local Feature Network (RLFN) (\citealt{li2023ntire})capable of super-resolving solar magnetic field images. RLFN is a novel Efficient Super-Resolution (ESR)(\citealt{kong2022residual}) scheme that uses three convolutional layers for residual local feature learning to simplify feature aggregation. This processing mechanism substantially reduces inference times while maintaining model capacity. For evaluation purposes, we compared the model-generated results with bicubic interpolation, SRCNN models and conducted a qualitative analysis of physical parameters between the generated magnetograms and corresponding SDO/HMI magnetograms. The structure of this paper is as follows: Section 2 introduces the data acquisition process and methods, as well as the neural network's architecture and training strategy. In Section 3, we evaluate the accuracy and performance of the trained models using a test dataset. Finally, Section 4 presents the conclusions, discussing the nature and future work on the subject.



\section{Data and Methods}
\label{sect:Obs}
\subsection{Data processing}

\begin{figure}[ht]
  \centering
  \includegraphics[width=0.80\textwidth]{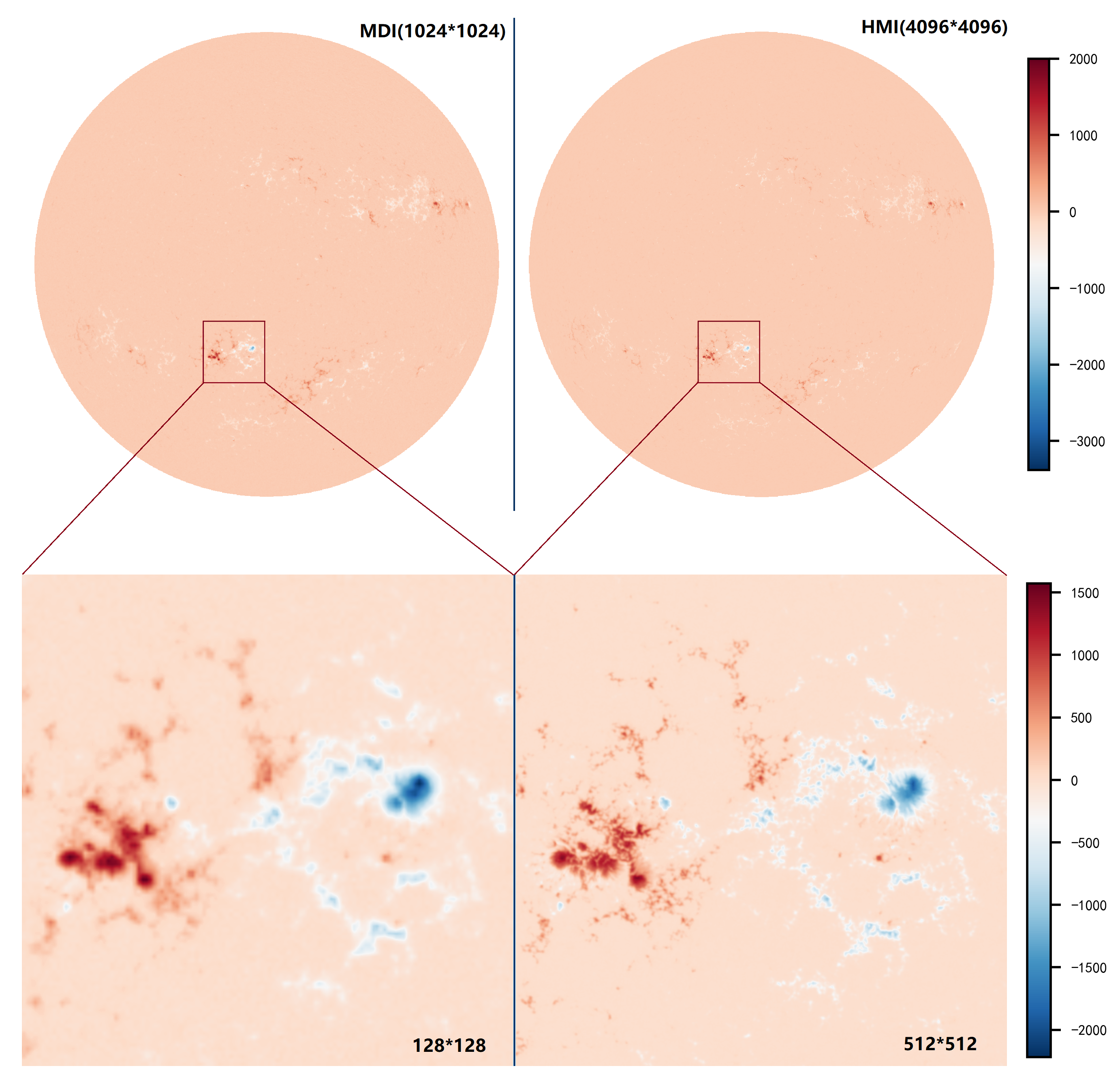}
  \caption{ Comparison of co-temporal, co-aligned magnetograms obtained by MDI (left), and HMI(right). Both images show the full solar disk, re-scaled as if they were observed from 1 Astronomical Unit, and plotted over the range of $\pm2000$ Gauss.}
  \label{fig:Fulldisk} 
\end{figure}

We have 20,150 sets of simultaneous full-disk observations from SOHO/MDI and SDO/HMI dating from April 8, 2010, to April 11, 2011, providing a rich dataset for model training. In this paper, data from April 2010 to February 2011 are allocated as training sets, data from March 2011 as validation sets, and data from April 2011 as test sets. Although neural networks can learn the systematic differences between instruments, random locations, rotations, and temporal differences between images can significantly hinder the network from learning mapping relationships. Therefore, preprocessing is required before training: 
\begin{enumerate}
\item pair MDI and HMI magnetograms with relatively close timing based on the T\_REC timestamp in the header files; 
\item use the SIFT algorithm(\citealp{alhwarin2008improved}) to register images between MDI and HMI full-disk magnetograms; 
\item randomly crop ten  $128 \times 128$  sub-blocks on each MDI full-disk magnetogram, with the variance of each sub-block being computed simultaneously. The sub-block with the highest variance is selected as the training data. A corresponding  $512 \times 512$  sub-block is then cropped from the HMI data at the equivalent location, followed by further alignment;
\item rotate these sub-blocks randomly at $0\degree$, $90\degree$, $180\degree$, and $270\degree$ and invert their polarity to implement data augmentation. 
\end{enumerate} 
Figure~\ref{fig:Fulldisk} showcases the preprocessed full-disk magnetograms.

\subsection{Methods}
Super-resolution image reconstruction (SRIR or SR) refers to the technique of converting existing low-resolution (LR) images into high-resolution (HR) images through signal and image processing methods, employing software algorithms(\citealt{freeman2000learning}). In astronomical data processing, SR implies utilizing a wealth of training data to learn the non-linear mapping relationship between LR and HR images, adding high-frequency content to observed astronomical images to compensate for quality degradation caused by astronomical telescope image processing(\citealt{yang2019deep}), and thereby acquiring higher-resolution astronomical images. Super-resolution is an ill-posed problem(\citealt{2019Single}) since multiple SR outputs can correspond to a single LR input. With the rapid advancement of deep learning and the emergence of high-performance GPUs, deep learning-based super-resolution reconstruction methods have become the main research direction(\citealt{dong2014learning}). CNNs and GANs represent two different deep learning models. GANs are primarily used for generating synthetic and fabricated data(\citealt{goodfellow2014generative},\citealt{creswell2018generative}), with studies by Sumiaya Rahman and Dou et al. focused on creating visually high-quality super-resolved magnetograms using GANs(\citealt{ledig2017photo}). CNNs, on the other hand, are mainly applied in image and vision tasks, offering greater physical interpretability(\citealt{lin2017does}).

\subsubsection{Modified Efficient Super-Resolution Model}
The field of single-image super-resolution is evolving rapidly with many eminent models emerging(\citealt{kim2016accurate},\citealt{lim2017enhanced},\citealt{yu2018wide}). General imagery has a more complex structure (edges, textures, and other semantic information) and therefore requires substantial computational resources for SR networks. In contrast, the super-resolution for magnetograms focuses more on fidelity(\citealt{baso2018enhancing}), preserving the constancy of magnetic field physical parameters(\citealt{guo2021nonlinear}) (like invariant magnetic flux) and deployment on devices with limited computing capabilities.
\\
RLFN(\citealt{kong2022residual}) is a novel Efficient Super-Resolution (ESR) model that employs three convolutional layers to perform residual local feature learning to streamline feature aggregation. This mechanism significantly reduces inference time while maintaining a moderate model size. Based on our preliminary research, we have gathered the following insights:
\begin{enumerate}[label=\textcircled{\arabic*}]
\item A larger receptive field and deeper network architecture contribute to enhanced fitting capabilities. As network depth increases, to counteract vanishing and exploding gradients, the introduction of residual structures becomes indispensable.
\item Positioning sub-pixel convolution at the end for image upscaling minimizes the introduction of artificial noise and allows feature computation at a lower resolution, thus significantly reducing computational demand.
\item Garnering more extensive feature maps prior to the activation layer and eliminating redundant activation and convolutional layers can decrease computational costs and stabilize the training process.
\end{enumerate}

\begin{figure}[ht]
\centering
\includegraphics[width=1.0\textwidth]{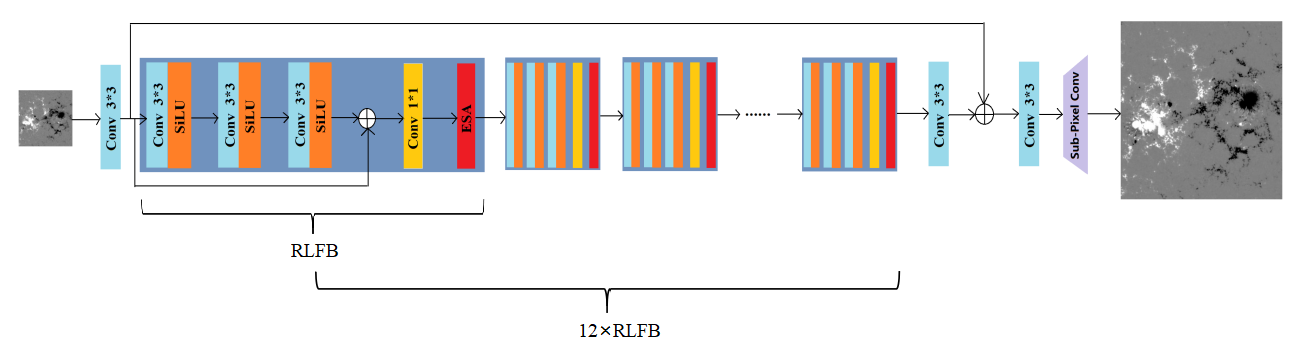}
\caption{The 100-layer deep convolutional neural network architecture of MESR.}
\label{fig:mesr_architecture}
\end{figure}

\begin{figure}[ht]
\centering
\includegraphics[width=0.5\textwidth]{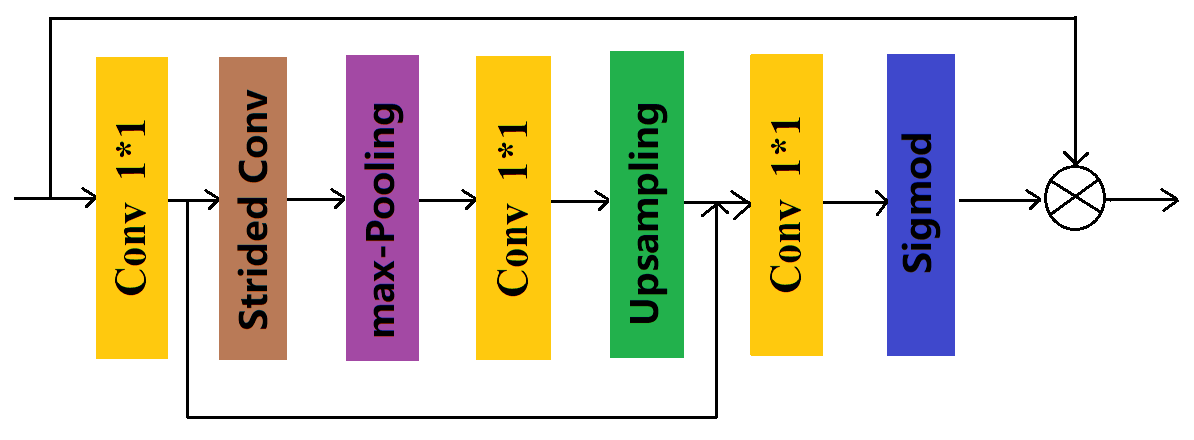}
\caption{The structure of the ESA module.}
\label{fig:esa_module}
\end{figure}

In this paper, we introduce an Efficient Super-Resolution model based on RLFN(\citealt{kong2022residual}), dubbed \textbf{MESR}, with a 100-layer deep convolutional neural network architecture as illustrated in Figure~\ref{fig:mesr_architecture}. The network comprises three main components: a shallow feature extraction layer (one $3 \times 3$ convolutional layer), a deep feature extraction layer (consisting of 12 stacked RLFBs), and an image reconstruction module (sub-pixel convolution layer). At its core lies the \textbf{RLFB}, proposed by Kong et al., where each block executes local feature extraction by stacking 3 \textit{Conv-SiLU} operations, uses a channel count of 64 to counteract performance degradation, which is followed by a $1 \times 1$ convolutional layer to reduce channel dimensions and connect with the \textbf{ESA} module, as depicted in Figure~\ref{fig:esa_module}. The \textbf{ESA} module initiates with a $1 \times 1$ convolution to lessen the channel dimensions of input features, then utilizes strided convolutions and max-pooling with a kernel size of $7 \times 7$ and stride of 3 to reduce spatial dimensions and boost efficiency. Followed by $3 \times 3$ convolutions for feature extraction and interpolation-based upsampling to restore spatial dimensions, it uses residual connections and processes the extracted features through a $1 \times 1$ convolutional layer to restore the channel count, culminating with a sigmoid-activated feature matrix that multiplies with the original input features.

\subsubsection{Loss Functions}
A loss function is a mathematical formula used to quantify the discrepancy between actual values and the predictions made by a model. In deep learning, loss functions are essential for evaluating network performance. The design of the loss function plays a pivotal role, influencing both the training speed and the distribution of data within the hidden layers. Selecting an appropriate loss function is vital for achieving desired network outcomes(\citealt{barron2019general}).
\\
During our experiments, although MSE loss provided visually satisfactory SR reconstruction results(\citealt{buhlmann2003boosting}), its sensitivity to larger errors often led to gradient explosion and over-smoothing compensation in the output. Thus, we experimented with different loss functions such as L1(\citealt{janocha2017loss}), SSIM, and PSNR to optimize the outcome. We observed that L1 loss excelled in both quantitative and visual assessments, equation 1:
\begin{equation}\label{eq1}
  \text{MAE(X,h)} = \frac{1}{m} \sum_{i=1}^{m} |h(x_i) - y_i|
\end{equation}
However, the downside is that the L1 loss is non-smooth and non-differentiable at zero,thus gradient descent cannot proceed at \(w=0\). Building upon this, in order to make the training more stable,we have explored a generalized loss function, known as the Charbonnier Loss, which is formulated as follows:
\begin{equation}\label{eq2}
  \text{Charbonnier Loss}(y, \hat{y})  = \frac{1}{N} \sum_{i=1}^{N} \rho(y - \hat{y})
\end{equation}
\begin{equation}\label{eq3}
  \rho(x) = \sqrt{x^2 + \epsilon^2}
\end{equation}
Compared to L1 loss, the curve of the Charbonnier Loss is smoother. Near-zero gradients are not too small due to the constant \(epsilon\), preventing gradient disappearance, while gradients for larger magnitudes are moderated by the square root, averting gradient explosion and handling outliers more effectively.

\section{Results and Discussion}
\label{soft}
  \label{fig:45compare}
\begin{figure}[htbp]
  \centering
  \includegraphics[width=0.9\textwidth]{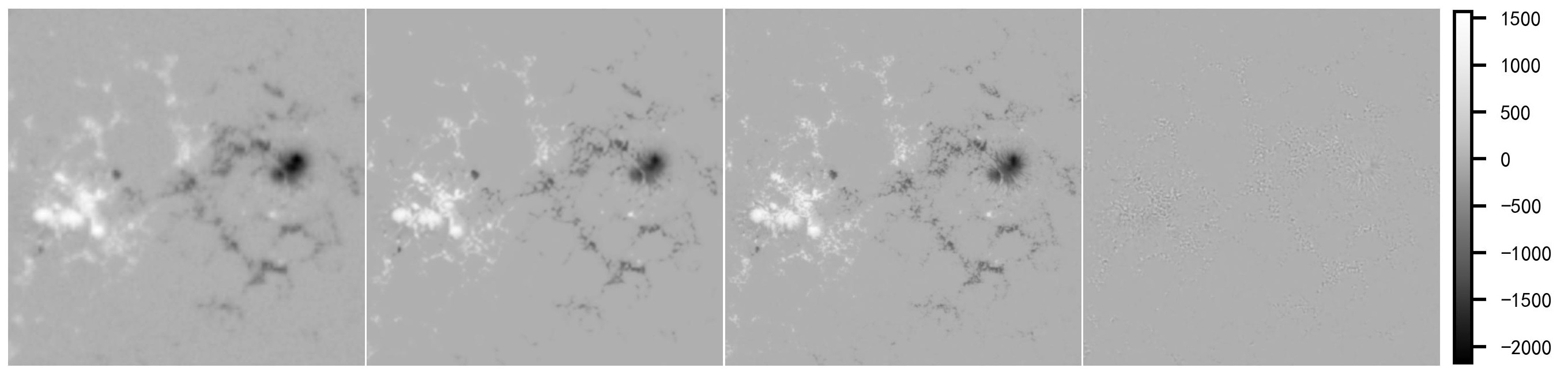}
  \caption{Super-resolution reconstruction of a local MDI magnetogram by MESR.From left to right : MDI LR ($128\times128$),MDI SR ($512\times512$),HMI original images($512\times512$), and residual graphs.}
  \label{fig:superres}
\end{figure}

The model training process involved cropping $128\times128$ LR MDI magnetograms and $512\times512$ HR HMI magnetograms to create data pairs, with a batch size of 32 and spanning 500 epochs, using the Adam optimizer. The initial learning rate was set at 0.001 and decreased over time. We utilized Charbonnier Loss to optimize our model, with a constant $\epsilon$ of 0.001. Figure~\ref{fig:superres} illustrates the super-resolution reconstruction of a local MDI magnetogram by MESR.Figure~\ref{fig:45compare} shows three scatter plots, from left to right, each representing the magnetograms reconstructed by the Bicubic, SRCNN, and MESR methods, respectively, compared with the scatter plots of the HMI high-resolution magnetogram. By comparison, it is observed that the points in the MESR scatter plot are closer to the diagonal line.

\begin{figure}[h]
  \centering
  \includegraphics[width=1.0\textwidth]{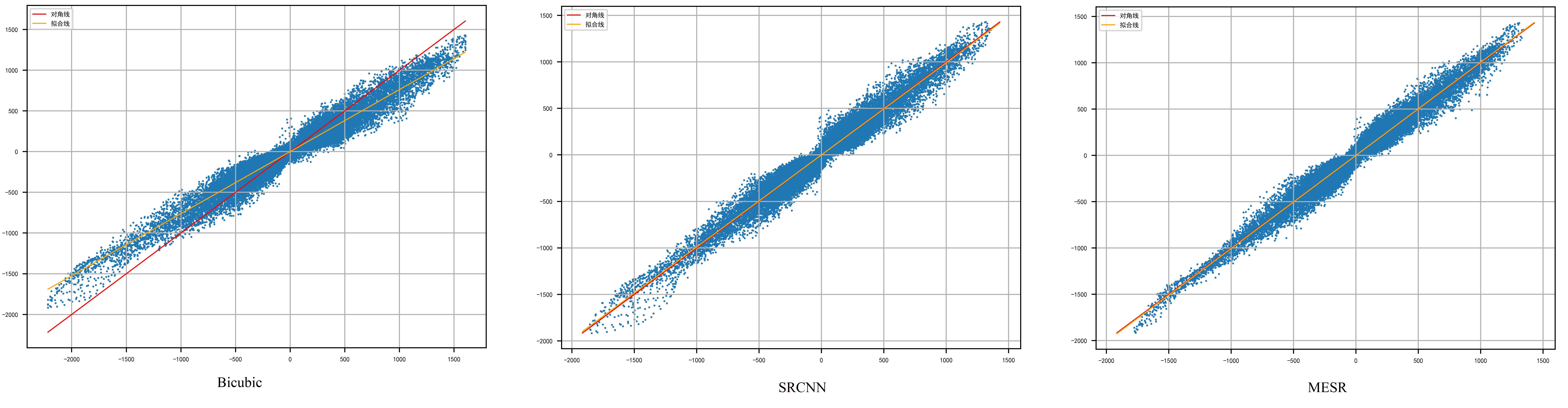}
  \caption{Scatter plots of the HMI HR and magnetograms generated by different models within the range of $-1500 \, \text{G}$ to 1500 G.}
  \label{fig:45compare}
\end{figure}

\begin{figure}[htbp]
  \centering
  \includegraphics[width=1.0\textwidth]{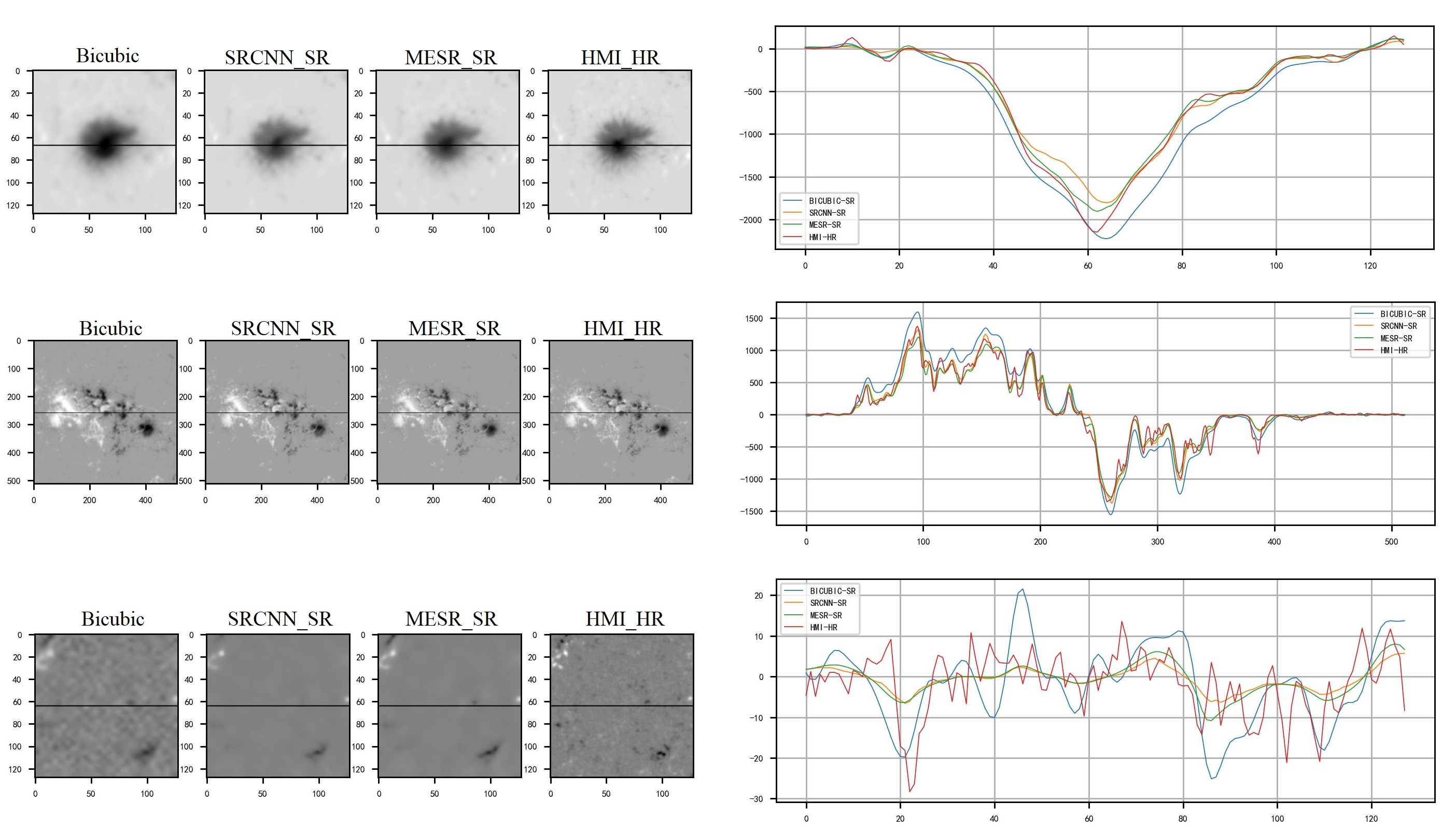}
  \caption{Variations in pixel grayscale values along a selected row in the reconstructed solar magnetograms by different models.The horizontal axis represents the lateral position (x-coordinate) on the solar magnetogram, while the vertical axis displays the corresponding pixel grayscale value.}
  \label{fig:solar_magnetograms}
\end{figure}

\begin{figure}[htbp]
  \centering
  \includegraphics[width=0.8\textwidth]{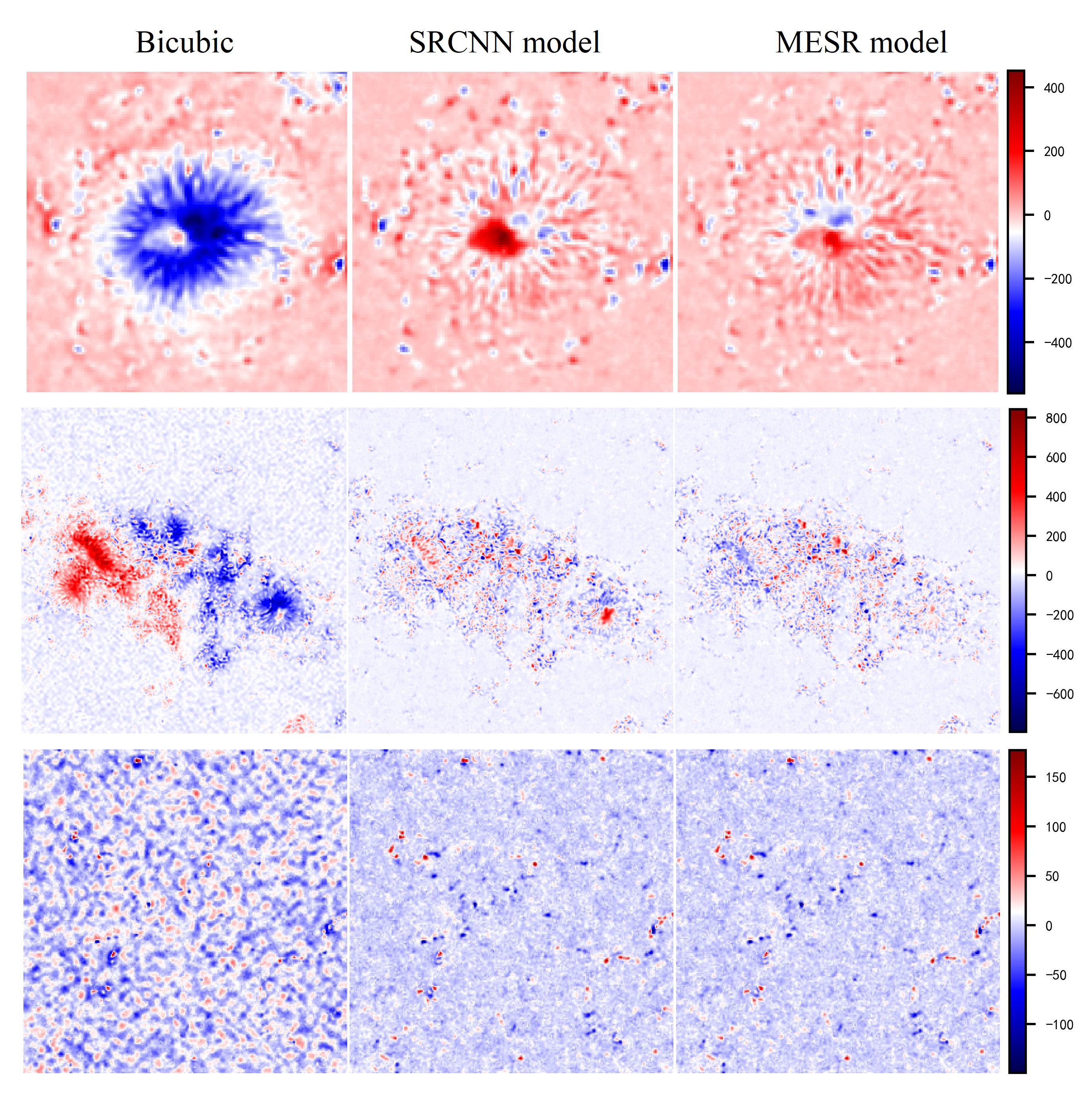}
  \caption{The absolute residual maps show the disparity between the reconstruction outcome and the HR magnetogram obtained from HMI (The formula for absolute error is given by $AE = SR - HR$.). Three sets of distinct testing data are displayed from top to bottom.}
  \label{fig:residual_maps}
\end{figure}

\begin{figure}[htbp]
  \centering
  \includegraphics[width=0.8\textwidth]{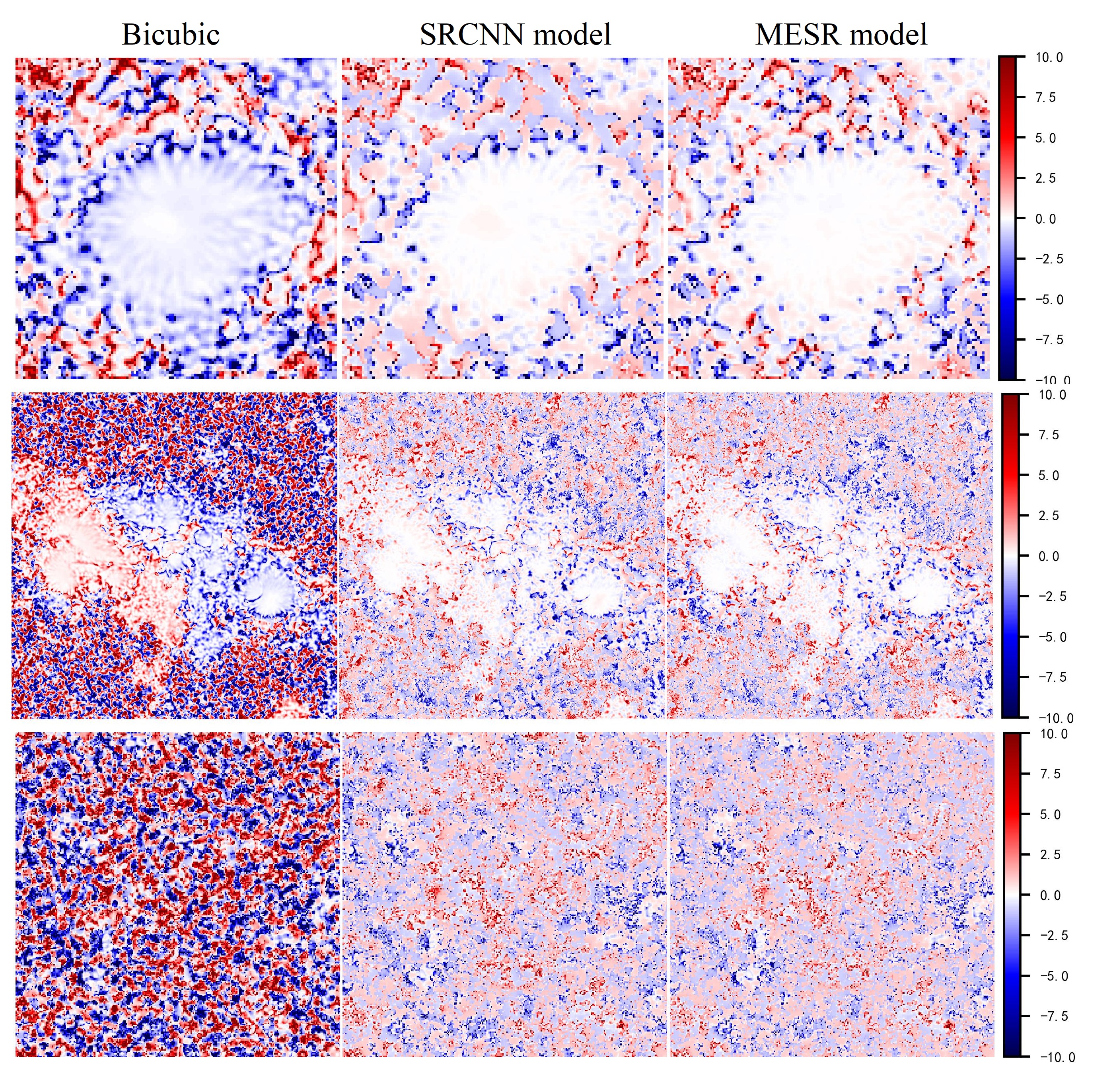}
  \caption{The relative residual maps (calculated by the formula $RE = \frac{AE}{\left| HR \right|}$).Three sets of distinct testing data are displayed from top to bottom.}
  \label{fig:Relative residual maps}
\end{figure}

Qualitative comparison: To evaluate the performance, the MESR model was employed on the test dataset and its results were visually compared with those obtained using bicubic interpolation and SRCNN, as shown in Figure~\ref{fig:solar_magnetograms}.Bicubic interpolation is the most fundamental method in image processing, and SRCNN represents the pioneering work of deep learning in the domain of image super-resolution reconstruction. This figure illustrates the variations in pixel grayscale values along a selected row in the reconstructed solar magnetograms by different models. The horizontal axis represents the horizontal position (x-coordinate) on the solar magnetogram, while the vertical axis indicates the corresponding pixel grayscale values. Such comparison allows for a straightforward analysis of the differences in how each model performs in reconstructing the fine details of the solar magnetic field.The absolute residual is shown in Figure~\ref{fig:residual_maps}. However, it is worth noting that in solar magnetic field data, the magnetic field strength can vary considerably between different regions (quiet and active regions). Consequently, Figure~\ref{fig:Relative residual maps} presents the relative residual map of the reconstruction results.The results indicate that MESR is more capable of reconstructing small-scale magnetic structures and reducing blur artifacts.

Quantitative evaluation: Crucially, the scientific value of super-resolution magnetograms cannot be solely affirmed by visual effects. For quantitative evaluation, we employ metrics such as Peak Signal-to-Noise Ratio (PSNR)(\citealt{janocha2017loss}), Structural Similarity Index (SSIM) (\citealt{hore2010image}), Correlation Coefficient (CC)(\citealt{asuero2006correlation}), and Root Mean Square Error (RMSE)(\citealt{chai2014root}). These assess our method's fidelity and linear correlations. PSNR measuring pixel-wise error, which can be used to measure the reconstruction quality of the model and reflect the fidelity of the generated image. The formula is as follows:
\begin{equation}\label{eq4}
  PSNR = 10 \cdot \log_{10} \left(\frac{MAX_I^2}{MSE}\right)= 20 \cdot \log_{10} \left(\frac{MAX_I}{\sqrt{MSE}}\right)
\end{equation}
In the formula, $i$ represents the number of test samples, $MAX_i$ denotes the maximum image grayscale range which is $4000G$, and the PSNR values typically vary from $30$ to $60$, with higher values being preferable.The Structural Similarity Index (SSIM) measures the similarity of images based on luminance, contrast, and structure. It uses the mean as an estimate of luminance, the standard deviation as an estimate of contrast, and the covariance as a measure of structural similarity. The formula is as follows:
\begin{equation}\label{eq5}
  \text{SSIM}(x, y) = \left[ \frac{(2 \mu_x \mu_y + c_1)(2 \sigma_{xy} + c_2)}{(\mu_x^2 + \mu_y^2 + c_1)(\sigma_x^2 + \sigma_y^2 + c_2)} \right]
\end{equation}
\begin{equation}\label{eq6}
  \text{l}(x, y) = \frac{2 \mu_x \mu_y + c_1}{\mu_x^2 + \mu_y^2 + c_1}
\end{equation}
\begin{equation}\label{eq7}
  c(x, y) = \frac{2\sigma_x\sigma_y + c_2}{\sigma_x^2 + \sigma_y^2 + c_2}
\end{equation}
\begin{equation}\label{eq8}
  s(x, y) = \frac{\sigma_{xy} + c_3}{\sigma_x \sigma_y + c_3}
\end{equation}
where, $\mu_x$, $\mu_y$, $\sigma_{xy}$, $\sigma_x^2$, $\sigma_y^2$ in the formula represent the mean of the target magnetic map, the mean of the reconstructed magnetic map, the covariance of the target and reconstructed magnetic maps, the variance of the target magnetic map, and the variance of the reconstructed magnetic map.$\mu_x$ represents the mean of the target magnetic map, $\mu_y$ represents the mean of the reconstructed magnetic map, $\sigma_{xy}$ represents the covariance of the target and reconstructed magnetic maps, $\sigma_x^2$ represents the variance of the target magnetic map, and $\sigma_y^2$ represents the variance of the reconstructed magnetic map. $c_1$ and $c_2$ are constants.SSIM is a value ranging from 0 to 1, where a value closer to 1 indicates a better similarity.
CC is the pixel-to-pixel Pearson correlation coefficient between the target magnetic map and the reconstructed magnetic map, reflecting the degree of linear correlation between the observed magnetic flux and the generated magnetic flux.
\begin{equation}\label{eq9}
  cc = \frac{\sum (I - I_{avg})(I_{SR} - I_{avg_{SR}})}{\sqrt{\sum (I - I_{avg})^2 \sum (I_{SR} - I_{avg_{SR}})^2}}
\end{equation}
Where $I$ and $I_{SR}$ respectively represent the actual measured magnetic flux and the reconstructed magnetic flux, $I_{avg}$ and $I_{avg_{SR}}$ represent the average magnetic flux of the actual measured data and the reconstructed data, respectively.
The final evaluation metric is the Root Mean Square Error (RMSE), which represents the mean square deviation between the target and generated magnetic maps. It reflects the deviation between the magnetic flux of the reconstruction result and the actual measured magnetic flux. The formula is given by:
\begin{equation}\label{eq10}
  RMSE = \sqrt{\frac{1}{N} \sum_{i=1}^{N}(I - I_{SR})^2}
\end{equation}

\begin{figure}[htbp]
  \centering
  \includegraphics[width=0.8\textwidth]{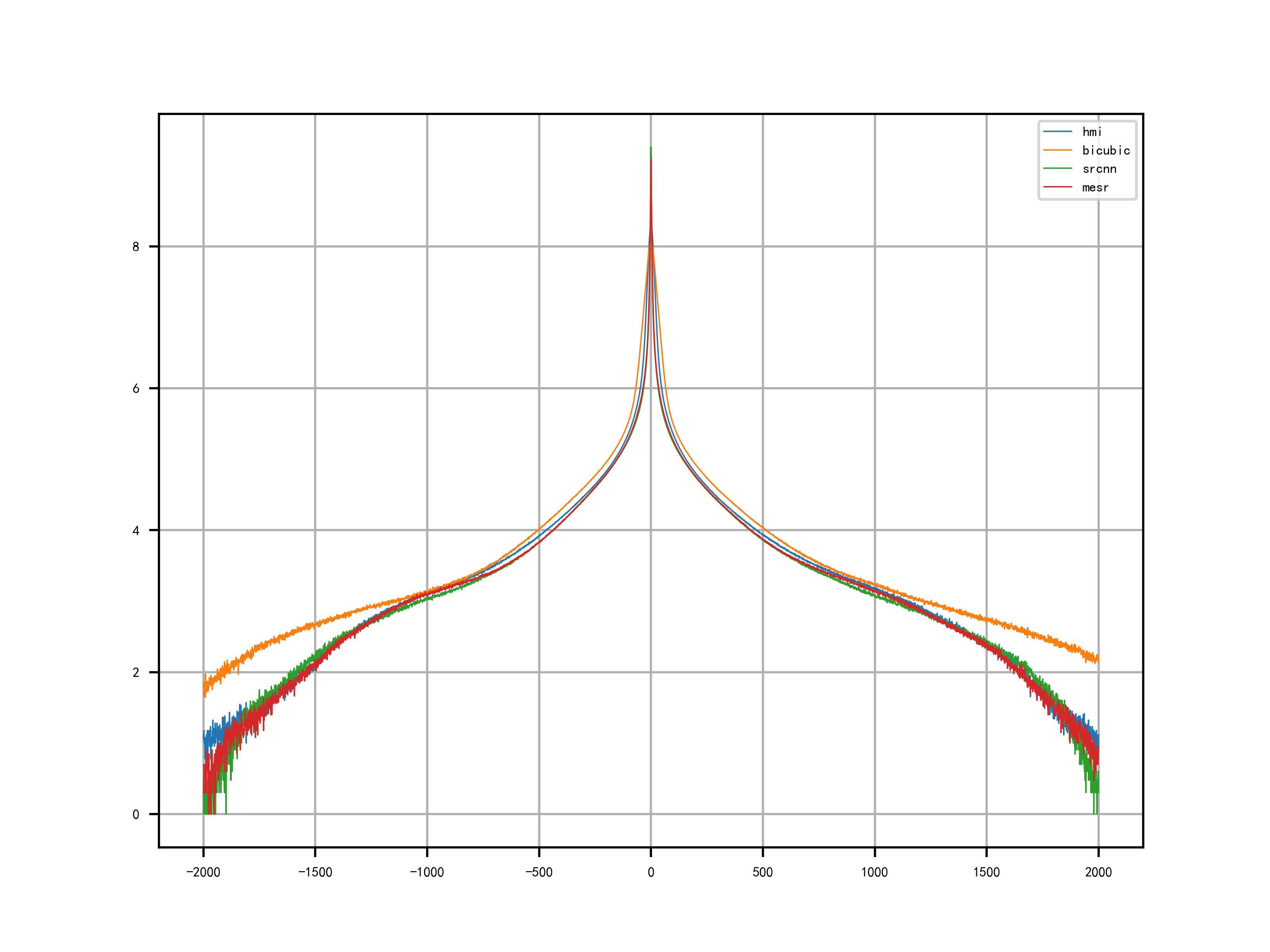}
  \caption{ Pixel statistical histogram of the full-disk test dataset reconstruction results.The horizontal axis represents the grayscale value range, while the vertical axis represents the logarithmically scaled count of pixels.}
  \label{fig:2D}
\end{figure}

Batch reconstruction on the test dataset was conducted using the MESR model, To evaluate the reconstruction quality of the testing dataset (MDI full-disk magnetogram), we can create a pixel histogram. This histogram will plot the distribution of pixel values for the reconstructed results of the entire testing dataset, providing insights into the dynamic range and potential biases in the reconstruction process.As shown in Figure~\ref{fig:2D}
And the quantitative evaluation of the statistical results is presented in Table~\ref{tab:methods_comparison}. Table~\ref{tab:methods_comparison} reports the average values of PSNR, SSIM, LPIPS, CC, and RMSE derived from the comparison between the resolution-enhanced full-disk magnetograms of SOHO/MDI and the full-disk magnetograms from SDO/HMI. The results clearly indicate that the MESR method significantly surpasses other approaches across all metrics, providing substantial evidence of its efficacy.

\begin{table}[htbp]
  \centering
  \caption{Image Processing Methods Comparison}
  \label{tab:methods_comparison}
  \begin{tabular}{lccccc}
  \hline
  Method                  & Pixel to Pixel (CC) & Peak S/N (dB) & SSIM   & RMSE (G)    \\
  \hline
  Bicubic Interpolation   & 0.828               & 46.779         & 0.954  & 18.506     \\
  SRCNN                   & 0.896               & 50.63          & 0.985  & 11.774     \\
  MESR                    & 0.911               & 51.112         & 0.987  & 11.181     \\
  \hline
  \end{tabular}
  \end{table}

\section{Conclusion and Future Work}
To thoroughly understand the solar activity cycle, high-quality data spanning time scales is required, and differences between measuring instruments hinder the integrated use of all available data. In this research, we applied deep learning to solar magnetic field imaging, creating a MESR network model capable of super-resolution reconstruction of SOHO/MDI images. It achieved reliable 4x super-resolution reconstructions for full-disk MDI magnetograms based on the high correlation coefficients (0.911) with the target HMI magnetograms. Elevated PSNR (51.112) and SSIM (0.987) values demonstrate the MESR model’s effectiveness. Visually, the MESR model surpassed bicubic interpolation and SRCNN in providing finer small-scale and textural details. Finally, 2D histograms and scatter plots in the MESR model displayed better pixel distribution and linear relationships than competing methods.
\\
There is, however, room for improvement in the high-frequency information fidelity in magnetic field images. Future work will focus on developing more viable loss functions, training strategies, and model architectures to tackle this issue and enhance results. Furthermore, interpretative and evaluative analyses based on solar physics indicators are planned, to bolster the credibility of data. Additionally, we intend to extend this methodology to other telescopes and instruments, laying the groundwork for multi-frame super-resolution reconstruction of all historical magnetogram data, paving the way for continuous, uniform, high-resolution magnetic field datasets crucial for space weather research.

\begin{acknowledgements}
  We appreciate the referee’s careful reading of the manuscript and many constructive comments, which helped greatly in improving the paper.This work was funded by the National Natural Science Foundation of China (NSFC, Nos. 12003068)and Yunnan Key Laboratory of Solar Physics and Space Science under the number 202205AG070009.
\end{acknowledgements}






%
\label{lastpage}

\bibliographystyle{raa} 
\bibliography{reference}

\end{document}